\PassOptionsToPackage{unicode}{hyperref}
\PassOptionsToPackage{hyphens}{url}
\PassOptionsToPackage{dvipsnames,svgnames,x11names}{xcolor}
\documentclass[
]{extarticle}

\usepackage{amsmath,amssymb}
\usepackage{iftex}
\ifPDFTeX
  \usepackage[T1]{fontenc}
  \usepackage[utf8]{inputenc}
  \usepackage{textcomp} 
\else 
  \usepackage{unicode-math}
  \defaultfontfeatures{Scale=MatchLowercase}
  \defaultfontfeatures[\rmfamily]{Ligatures=TeX,Scale=1}
\fi
\usepackage{lmodern}
\ifPDFTeX\else  
\fi
\IfFileExists{upquote.sty}{\usepackage{upquote}}{}
\IfFileExists{microtype.sty}{
  \usepackage[]{microtype}
  \UseMicrotypeSet[protrusion]{basicmath} 
}{}
\makeatletter
\@ifundefined{KOMAClassName}{
  \IfFileExists{parskip.sty}{%
    \usepackage{parskip}
  }{
    \setlength{\parindent}{0pt}
    \setlength{\parskip}{6pt plus 2pt minus 1pt}}
}{
  \KOMAoptions{parskip=half}}
\makeatother
\usepackage{xcolor}
\makeatletter
\ifx\paragraph\undefined\else
  \let\oldparagraph\paragraph
  \renewcommand{\paragraph}{
    \@ifstar
      \xxxParagraphStar
      \xxxParagraphNoStar
  }
  \newcommand{\xxxParagraphStar}[1]{\oldparagraph*{#1}\mbox{}}
  \newcommand{\xxxParagraphNoStar}[1]{\oldparagraph{#1}\mbox{}}
\fi
\ifx\subparagraph\undefined\else
  \let\oldsubparagraph\subparagraph
  \renewcommand{\subparagraph}{
    \@ifstar
      \xxxSubParagraphStar
      \xxxSubParagraphNoStar
  }
  \newcommand{\xxxSubParagraphStar}[1]{\oldsubparagraph*{#1}\mbox{}}
  \newcommand{\xxxSubParagraphNoStar}[1]{\oldsubparagraph{#1}\mbox{}}
\fi
\makeatother

\usepackage{longtable,booktabs,array}
\usepackage{calc} 
\usepackage{etoolbox}
\makeatletter
\patchcmd\longtable{\par}{\if@noskipsec\mbox{}\fi\par}{}{}
\makeatother
\IfFileExists{footnotehyper.sty}{\usepackage{footnotehyper}}{\usepackage{footnote}}
\makesavenoteenv{longtable}
\usepackage{graphicx}
\makeatletter
\def\maxwidth{\ifdim\Gin@nat@width>\linewidth\linewidth\else\Gin@nat@width\fi}
\def\maxheight{\ifdim\Gin@nat@height>\textheight\textheight\else\Gin@nat@height\fi}
\makeatother
\setkeys{Gin}{width=\maxwidth,height=\maxheight,keepaspectratio}
\makeatletter
\def\fps@figure{htbp}
\makeatother
\NewDocumentCommand\citeproctext{}{}

\makeatletter
 \let\@cite@ofmt\@firstofone
 \def\@biblabel#1{}
 \def\@cite#1#2{{#1\if@tempswa , #2\fi}}
\makeatother
\newlength{\cslhangindent}
\newlength{\csllabelwidth}
\newenvironment{CSLReferences}[2] 
 {\begin{list}{}{%
  \setlength{\itemindent}{0pt}
  \setlength{\leftmargin}{0pt}
  \setlength{\parsep}{0pt}
  \ifodd #1
   \setlength{\leftmargin}{\cslhangindent}
   \setlength{\itemindent}{-1\cslhangindent}
  \fi
  \setlength{\itemsep}{#2\baselineskip}}}
 {\end{list}}
\usepackage{calc}

\makeatletter
\@ifpackageloaded{caption}{}{\usepackage{caption}}
\AtBeginDocument{%
\ifdefined\contentsname
  \renewcommand*\contentsname{Table of contents}
\else
  \newcommand\contentsname{Table of contents}
\fi
\ifdefined\listfigurename
  \renewcommand*\listfigurename{List of Figures}
\else
  \newcommand\listfigurename{List of Figures}
\fi
\ifdefined\listtablename
  \renewcommand*\listtablename{List of Tables}
\else
  \newcommand\listtablename{List of Tables}
\fi
\ifdefined\figurename
  \renewcommand*\figurename{Figure}
\else
  \newcommand\figurename{Figure}
\fi
\ifdefined\tablename
  \renewcommand*\tablename{Table}
\else
  \newcommand\tablename{Table}
\fi
}
\@ifpackageloaded{float}{}{\usepackage{float}}
\floatstyle{ruled}
\@ifundefined{c@chapter}{\newfloat{codelisting}{h}{lop}}{\newfloat{codelisting}{h}{lop}[chapter]}
\floatname{codelisting}{Listing}

\makeatother
\makeatletter
\@ifpackageloaded{caption}{}{\usepackage{caption}}
\@ifpackageloaded{subcaption}{}{\usepackage{subcaption}}
\makeatother

\ifLuaTeX
  \usepackage{selnolig}  
\fi
\usepackage{bookmark}

\IfFileExists{xurl.sty}{\usepackage{xurl}}{} 
\hypersetup{
  pdftitle={EphemerisSources.jl: Idiomatic Ephemeris Sourcing and Parsing in Julia},
  colorlinks=true,
  linkcolor={blue},
  filecolor={Maroon},
  citecolor={Blue},
  urlcolor={Blue},
  pdfcreator={LaTeX via pandoc}}

\title{EphemerisSources.jl: Idiomatic Ephemeris Sourcing and Parsing in
Julia}
\author{Joseph D. Carpinelli}
\date{2024-04-21}

\begin{document}
\maketitle

\section{Summary}\label{summary}

Students and professionals in astronomy, astrodynamics, astrophysics,
and other related fields often download and parse data about objects in
our solar system --- ephemeris data --- from two major providers: JPL's
publicly-available
\href{https://naif.jpl.nasa.gov/pub/naif/generic_kernels/}{Generic SPICE
Kernels} and JPL's \href{https://ssd.jpl.nasa.gov/horizons/}{Horizons
platform}. SPICE kernels are typically read through the SPICE Toolkit,
which is available in a variety of programming languages, including the
C Programming Language with \texttt{CSPICE} (Acton 1996). The Julia
packages
\href{https://github.com/JuliaBinaryWrappers/CSPICE_jll.jl}{\texttt{CSPICE\_jll.jl}}
and \href{https://github.com/JuliaAstro/SPICE.jl}{\texttt{SPICE.jl}}
expose many \texttt{CSPICE} functions through Julia functions. Julia
users can load and interact with SPICE kernels through methods such as
\texttt{SPICE.furnsh} and \texttt{SPICE.spkez}. Horizons provides data
through a variety of methods, including email, command-line, graphical
web interfaces, and a
\href{https://ssd-api.jpl.nasa.gov/doc/horizons.html}{REST API}
(Giorgini 2015).

This paper introduces several packages --- \texttt{SPICEKernels.jl},
\texttt{SPICEBodies.jl}, \texttt{HorizonsAPI.jl} and
\texttt{HorizonsEphemeris.jl} --- which allow users to download and
process Cartesian state vector data idiomatically, all from within
Julia. While ephemeris data comes in many forms, including observer
tables, osculating orbital elements, and binary formats, these packages
currently target Cartesian state vector (position and velocity)
ephemeris data. Through the use of these packages, users can share
replicable code which automatically fetches data from publicly-available
ephemeris sources, as opposed to manually including ephemeris data files
with their source code distribution.

\section{Statement of Need}\label{statement-of-need}

While astronomers, astrodynamicists, and other ephemeris users have the
tools they need to fetch and parse position and velocity data from
multiple sources within Julia, they do not have the tools to do so
\emph{simply} or \emph{idiomatically}. Horizons' ephemeris data is
distributed in plain text with surrounding metadata, and manual parsing
is required for users to programmatically use the fetched ephemeris
data. Generic SPICE kernels are freely available and can be used with
\texttt{CSPICE} (and wrapper libraries) for kernel inspection and data
retrieval, but new users and students may find the required workflows
unfamiliar. The packages presented in this paper may be used by students
and professionals to idiomatically inspect and use Cartesian state
vector ephemeris data, without prior knowledge of SPICE Toolkit
utilities or REST APIs.

\subsection{JPL Horizons}\label{jpl-horizons}

The two Horizons-related packages presented in this paper ---
\href{https://github.com/JuliaAstro/EphemerisSources.jl/tree/main/lib/HorizonsAPI}{\texttt{HorizonsAPI.jl}}
and
\href{https://github.com/JuliaAstro/EphemerisSources.jl/tree/main/lib/HorizonsEphemeris}{\texttt{HorizonsEphemeris.jl}}
--- are respectively the first Julia packages to precisely match the
REST API with tab-completion through \emph{static keyword
arguments}\footnote{The code required to support static keyword
  arguments was provided by Joseph Wilson, as described in the
  \textbf{Acknowledgements} section.}, and the first to offer automatic
response parsing into \texttt{NamedTuple} types. The \texttt{NamedTuple}
output of \texttt{HorizonsEphemeris.ephemeris}, the top-level method for
fetching Cartesian state vectors from the Horizons platform, allows for
easy plotting, CSV file-saving, and \texttt{DataFrame} construction.
Both \texttt{HorizonsAPI.jl} and \texttt{HorizonsEphemeris.jl} offer
users a simple, repeatable way to query and parse Horizons state vector
data. Parsing for other ephemeris types, including observer tables and
osculating orbital element tables, are desired features but are not yet
implemented. For sending Horizons requests for these other ephemeris
types, use \texttt{HorizonsAPI} methods to manually construct each
request, or see
\href{https://github.com/PerezHz/Horizons.jl}{\texttt{Horizons.jl}}.

\subsection{JPL SPICE}\label{jpl-spice}

The two SPICE-related packages presented in this paper ---
\href{https://github.com/JuliaAstro/EphemerisSources.jl/tree/main/lib/SPICEKernels}{\texttt{SPICEKernels.jl}},
and
\href{https://github.com/JuliaAstro/EphemerisSources.jl/tree/main/lib/SPICEBodies}{\texttt{SPICEBodies.jl}}
--- provide idiomatic kernel fetching, inspection, and caching from
within Julia. Previously, Julia users interacted with SPICE kernels by
manually downloading publicly-available
\href{https://naif.jpl.nasa.gov/pub/naif/generic_kernels/}{generic
kernels} and parsing the data using \texttt{SPICE.jl}, or another
ephemeris parsing source. This workflow requires that users know how to
find the appropriate generic kernels for their chosen application, and
that they know how to use CSPICE functions to retrieve their desired
data. \texttt{SPICEKernels.jl} and \texttt{SPICEBodies.jl} offer
idiomatic interfaces to ephemeris fetching and parsing respectively. The
\texttt{SPICEKernels.jl} project uses continuous integration to fetch
and parse publicly-available
\href{https://naif.jpl.nasa.gov/pub/naif/generic_kernels}{kernels} and
expose each kernel as a variable in a new package release. SPICE Toolkit
executables, provided by
\href{https://github.com/JuliaAstro/SPICE.jl/tree/main/lib/SPICEApplications}{\texttt{SPICEApplications.jl}},
are used to fill each corresponding kernel variable's docstring with a
description of the kernel's contents. \texttt{SPICEKernels.jl} users can
utilize tab-completion and Julia's built-in documentation tools to help
select the most appropriate generic kernel for their application. Once
each kernel is loaded into the SPICE kernel pool with \texttt{SPICE.jl},
users can use \texttt{SPICEBodies.jl} to idiomatically fetch state
vector data at provided epochs.

\section{Usage}\label{usage}

For detailed usage examples, consult the common
\href{https://juliaastro.org/EphemerisSources.jl}{documentation site}.

\section{External Packages}\label{external-packages}

The packages presented in this paper which interact with the SPICE
Toolkit require users to use
\href{https://github.com/JuliaAstro/SPICE.jl}{\texttt{SPICE.jl}}, or
another SPICE-compatible kernel loading tool. \texttt{SPICEBodies.jl}
uses the kernel cache that is created with \texttt{SPICE.furnsh}.
Support for other SPICE kernel management packages, such as
\href{https://github.com/JuliaSpaceMissionDesign/Ephemerides.jl}{\texttt{Ephemerides.jl}},
may be added in the future. Support for \texttt{Ephemerides.jl} is
particularly desirable, as it enables fetching kernel data in
multi-threaded contexts. In addition to the packages in this paper which
interface with the JPL Horizons ephemeris platform, the
\href{https://github.com/PerezHz/Horizons.jl}{\texttt{Horizons.jl}}
package offers simplified interfaces for constructing and sending
queries to the JPL Horizons REST API. The \texttt{Horizons.jl} package
provides support for all Horizons query types, as does
\texttt{HorizonsAPI.jl}. As stated previously,
\texttt{HorizonsEphemeris.jl} currently only supports parsing for
Cartesian state vector data.

\section{Acknowledgements}\label{sec-acknowledgements}

Joseph Wilson (user \texttt{@jollywatt} on Julia's
\href{https://discourse.julialang.org/u/Jollywatt/summary}{Discourse})
provided incredibly helpful
\href{https://discourse.julialang.org/t/unpack-namedtuple-into-a-function-definition/97500}{guidance
and code} to support static keyword arguments. This contribution
substantially improved the usability of \texttt{HorizonsAPI.jl}.

\section{Disclaimers}\label{disclaimers}

The software developed in this paper, and the paper itself, was written
by the author in a personal capacity. This work does not reflect the
views of any organization, employer, or entity, except for the author as
an individual.

\section*{References}\label{references}
\addcontentsline{toc}{section}{References}

\phantomsection\label{refs}

\end{document}